\documentstyle[12pt,aps]{revtex}
\begin{document}
\draft
\author{Yishi Duan$^1$, Libin Fu$^{2,}$\thanks{%
Corresponding author. Email: lbfu@263.net } and Xing Liu$^1$}
\address{$^1$ Physics Department, Lanzhou University, Lanzhou 730000, P.R. China\\
$^2$ LCP, Institute of Applied Physics and Computational Mathematics, \\
P.O. Box 8009(26), Beijing 100088, P.R. China}
\title{The second Chern class in Spinning System}
\date{\today }
\maketitle

\begin{abstract}
\begin{center}
{\bf Abstract}
\end{center}

Topological property in a spinning system should be directly associated with
its wavefunction. A complete decomposition formula of $SU(2)$ gauge
potential in terms of spinning wavefunction is established rigorously. Based
on the $\phi $-mapping theory and this formula, one proves that the second
Chern class is inherent in the spinning system. It is showed that this
topological invariant is only determined by the Hopf index and Brouwer
degree of the spinning wavefunction.
\end{abstract}

\newpage

\section{introduction}

Topology now becomes absolutely necessary in physics\cite{t1,jmp28,t2,inden}%
. The $\phi $-mapping theory and the gauge potential decomposition theory%
\cite{D1,D2} are found to be significant in exhibiting the topological
structure of physics systems\cite{D3,D4,D5,Dz}. Topological properties of a
quantum system should be directly associated with its wavefunction. In our
previous paper hep-th%
\mbox{$\backslash$}
9908168, based on $\phi $-mapping theory and gauge potential decomposition
theory, we reveal the inner relation between the topological property of
Schr\"odinger system and the intrinsic properties of its wavefunction. We
point out that a topological invariant, the first Chern class, is inherent
in the Schr\"odinger system, which is only associated with the wave
function. This intrinsic relation is the topological source of the inner
structure of London equations in superconductor\cite{Dz}.

In this paper, we extend our work to spinning system. Based on the gauge
potential decomposition theory, we investigate the $SU(2)$ gauge potential
in terms of the spinning wavefunction. The complete decomposition formula of
the $SU(2)$ gauge potential is obtained rigorously. By making use of this
result, the second Chern class is studied. One proves that the second Chern
class is exclusively labeled by the topological index of the spinning
wavefunction without using any particular models or hypotheses. In other
words, one shows that the second Chern class is inherent in spinning system.
This relation between Chern class and wavefunction is intrinsic property of
quantum system, and gives a basic concept in topological quantum mechanics.

\section{$SU(2)$ gauge potential in terms of spinning wavefunction}

Considering a spinning wavefunction $\psi =\left( 
\begin{array}{c}
\psi ^1 \\ 
\psi ^2 
\end{array}
\right) ,$ we have the convariant derivative denoted as 
\begin{equation}
\label{cond}D\psi =d\psi -A\psi , 
\end{equation}
where $A=\frac i2A^a\sigma ^a$ $(a=1,2,3)$ is a $SU(2)$ gauge potential, and 
$\sigma ^a$ is the Pauli Matrices. Its complex conjugate is given by 
\begin{equation}
\label{ccond}D\psi ^{+}=d\psi ^{+}+\psi ^{+}A, 
\end{equation}
in which $\psi ^{+}=\left( 
\begin{array}{cc}
\psi ^{1*} & \psi ^{2*} 
\end{array}
\right) .$ By making use of the equation%
$$
\sigma ^a\sigma ^b+\sigma ^b\sigma ^a=2I\delta ^{ab}, 
$$
one can prove that 
\begin{equation}
\label{fulld}A^a=-\frac i{\psi ^{+}\psi }[(\psi ^{+}\sigma ^ad\psi -d\psi
^{+}\sigma ^a\psi )-(\psi ^{+}\sigma ^aD\psi -D\psi ^{+}\sigma ^a\psi )]. 
\end{equation}
From this formula, it can be proved that 
\begin{equation}
\label{sa}A^a=-\frac i{\psi ^{+}\psi }Tr(d\psi \psi ^{+}\sigma ^a-\psi d\psi
^{+}\sigma ^a)+\frac i{\psi ^{+}\psi }Tr(D\psi \psi ^{+}\sigma ^a-\psi D\psi
^{+}\sigma ^a). 
\end{equation}
Using the properties of Pauli matrices and considering $Tr(A)=0,$ we can
obtain that 
$$
A=\frac 1{\psi ^{+}\psi }[(d\psi \psi ^{+}-\psi d\psi ^{+})-Tr(d\psi \psi
^{+}-\psi d\psi ^{+})I] 
$$
\begin{equation}
\label{aall}-\frac 1{\psi ^{+}\psi }[(D\psi \psi ^{+}-\psi D\psi
^{+})-Tr(D\psi \psi ^{+}-\psi D\psi ^{+})I]. 
\end{equation}
From (\ref{aall}), it is easy to prove that $A$ satisfies the gauge
transformation 
$$
A^{^{\prime }}=SAS^{-1}+dSS^{-1}, 
$$
under the $SU(2)$ transformation 
$$
\psi ^{^{\prime }}=S\psi ,\quad \psi ^{^{\prime }+}=\psi ^{+}S^{+}. 
$$
Here, we know that $S^{+}=S^{-1}.$

The main feature of the decomposition theory of the gauge potential is that
the gauge potential $A$ can be generally decomposed as\cite{D1,D2,fujmp}%
$$
A=a+b, 
$$
where $a$ and $b$ are required to satisfy the following respective
transformation 
\begin{equation}
\label{aaa}a^{\prime }=SaS^{-1}+dSS^{-1},
\end{equation}
\begin{equation}
\label{bbb}b=SbS^{-1}.
\end{equation}
Investigating Eq. (\ref{aall}), one finds that the $SU(2)$ gauge potential
can be decomposed as 
\begin{equation}
\label{a}a=\frac 1{\psi ^{+}\psi }[(d\psi \psi ^{+}-\psi d\psi
^{+})-Tr(d\psi \psi ^{+}-\psi d\psi ^{+})I],
\end{equation}
and 
\begin{equation}
\label{b}b=-\frac 1{\psi ^{+}\psi }[(D\psi \psi ^{+}-\psi D\psi
^{+})-Tr(D\psi \psi ^{+}-\psi D\psi ^{+})I],
\end{equation}
which satisfy the Eqs. (\ref{aaa}) and (\ref{bbb}) respectively. We know
that the topological property is independent with the choice of the gauge
potential\cite{inden}. So, we can take $A$ as 
\begin{equation}
\label{amatric}A=\frac 1{\psi ^{+}\psi }[(d\psi \psi ^{+}-\psi d\psi
^{+})-Tr(d\psi \psi ^{+}-\psi d\psi ^{+})I].
\end{equation}
One can regard it as a special gauge. In fact, this choice is equal to the
common condition, $D\psi =0,$ which is often used in studying spinning
problem.

On the other hand, the spinning wavefunction can be denoted as%
$$
\psi =\left( 
\begin{array}{c}
\phi ^1+i\phi ^2 \\ 
\phi ^3+i\phi ^4 
\end{array}
\right) , 
$$
where $\phi ^a(a=1,2,3,4)$ are real function. One can regard the spinning
wavefunction as the complex representation of a vector field $\vec \phi
=(\phi ^1,\phi ^2,\phi ^3,\phi ^4)$. Let us define a unit vector $\vec n$ as 
\begin{equation}
\label{nm}n^a=\frac{\phi ^a}{||\phi ||},\quad ||\phi ||=(\phi ^a\phi ^a)^{%
\frac 12}=(\psi ^{+}\psi )^{\frac 12}, 
\end{equation}
and 
\begin{equation}
\label{na1}n^an^a=1. 
\end{equation}
If we denote that%
$$
N=\left( 
\begin{array}{c}
n^1+in^2 \\ 
n^3+in^4 
\end{array}
\right) ,\quad N^{+}=\left( 
\begin{array}{cc}
n^1-in^2 & n^3-in^4 
\end{array}
\right) , 
$$
then one has 
\begin{equation}
\label{na}N^{+}N=n^an^a=1 
\end{equation}
The expression (\ref{amatric}) can be rewrote as 
\begin{equation}
\label{am2}A=(NdN^{+}-dNN^{+})-Tr(NdN^{+}-dNN^{+})I. 
\end{equation}

\section{The second Chern class inherits in spinning system}

We know that the unit vector $\vec n$ can be expressed in terms of Clifford
algebra\cite{jmp22,jmp19} as 
\begin{equation}
\label{n}n=n^As_A,\qquad A=0,1,2,3. 
\end{equation}
with $n^0=n^4,$ and where 
\begin{equation}
\label{ss}s=(I,i\vec \sigma ),\qquad s^{\dagger }=(I,-i\vec \sigma ). 
\end{equation}
Then one can rewrite (\ref{na1}) as 
\begin{equation}
\label{nomal2}nn^{\dagger }=1. 
\end{equation}

If we choose the spinor as $\psi =\left( 
\begin{array}{c}
\phi ^1-i\phi ^2 \\ 
-\phi ^3-i\phi ^4 
\end{array}
\right) $, it can be proved that the gauge potential $A$ given in (\ref{am2}%
) has the expression as 
\begin{equation}
\label{result}A=dnn^{\dagger }\ . 
\end{equation}
It is interesting that this result is the same as the decomposition formula
of the $SU(2)$ gauge potential in terms of a four-dimensional vector field
which we have obtained in \cite{fujmp}.

One knows that the second Chern class can be wrote in terms of Chern-Simon%
\cite{inden,jmp25} 
\begin{equation}
\label{ch2a}C_2(P)=\frac 1{8\pi ^2}d\Omega , 
\end{equation}
and 
\begin{equation}
\Omega =\frac 1{8\pi ^2}Tr\left( A\wedge dA-\frac 23A\wedge A\wedge A\right) 
\end{equation}
which is known as Chern-Simon form\cite{jmp27}.

Substituting (\ref{result}) into (\ref{ch2a}) and considering (\ref{nomal2}%
), we obtain 
\begin{equation}
C_2(P)=\frac 1{24\pi ^2}Tr(dn\wedge dn^{\dagger }\wedge dn\wedge dn^{\dagger
}). 
\end{equation}
From (\ref{n}), one gets 
$$
C_2(P)=\frac 1{24\pi ^2}\in ^{\mu \nu \lambda \rho }\partial _\mu
n^A\partial _\nu n^B\partial _\lambda n^C\partial \rho n^DTr\left(
s_As_B^{\dagger }s_Cs_D^{\dagger }\right) dx^4 
$$
\begin{equation}
\label{tmrs}=\frac 1{12\pi ^2}\in ^{\mu \nu \lambda \rho }\in
_{ABCD}\partial _\mu n^A\partial _\nu n^B\partial _\lambda n^C\partial _\rho
n^Ddx^4. 
\end{equation}
By substituting (\ref{nm}) into (\ref{tmrs}), and considering 
\begin{equation}
dn^A=\frac{d\phi ^A}{||\phi ||}+\phi ^Ad\left( \frac 1{||\phi ||}\right) , 
\end{equation}
we have 
\begin{equation}
C_2(P)=-\frac 1{4\pi ^2}\frac{\partial ^2}{\partial \phi ^A\partial \phi ^A}%
{1 \overwithdelims() ||\phi ||^2}
D\left( \phi /x\right) d^4x, 
\end{equation}
where $D(\phi /x)$ is the Jacobian defined as 
\begin{equation}
\in ^{ABCD}D\left( \phi /x\right) =\in ^{\mu \nu \lambda \rho }\partial _\mu
\phi ^A\partial \nu \phi ^B\partial _\lambda \phi ^C\partial _\rho \phi ^D. 
\end{equation}
By means of the general Green function formula 
\begin{equation}
\frac{\partial ^2}{\partial \phi ^A\partial \phi ^A}\left( \frac 1{||\phi
||^2}\right) =-4\pi \delta ^4\left( \vec \phi \right) , 
\end{equation}
we get 
\begin{equation}
C_2(P)=\delta ^4\left( \vec \phi \right) D\left( \phi /x\right) d^4x. 
\end{equation}

Suppose $\phi ^A(x)\ (A=0,1,2,3)$ have $m$ isolated zeros at $x_\mu =z_\mu
^i\ (i=1,2,\cdot \cdot \cdot ,m)$ , according to the $\delta $-Function
theory\cite{jmp28}, $\delta (\vec \phi )$ can be expressed by 
\begin{equation}
\label{by}\delta (\vec \phi )=\sum_{i=1}^m\frac{\beta _i\delta (\vec x-\vec z%
_i)}{|D(\phi /x)|_{\vec x=\vec z_i}}, 
\end{equation}
and one then obtains 
\begin{equation}
C_2(P)=\sum_{i=1}^m\eta _i\beta _i\delta ^4\left( x-z_i\right) d^4x, 
\end{equation}
where $\beta _i$ is a positive integer (the Hopf index of the $i$th zeros )
and $\eta _i$ is the Brouwer degree\cite{jmp29}: 
\begin{equation}
\eta _i=\frac{D\left( \phi /x\right) }{|D\left( \phi /x\right) |}%
=sgn[D\left( \phi /x\right) ]|_{x=z_i}=\pm 1. 
\end{equation}
The meaning of the Hopf index $\beta _i$ is that the vector field function $%
\vec \phi $ covers the corresponding region $\beta _i$ times while $\vec x$
covers the region neighborhood of zero $z_i$ once. From above discussion,
the Chern density $\rho (x)$ is defined as : 
\begin{equation}
\label{roi}\rho (x)=\sum_{i=1}^m\eta _i\beta _i\delta ^4\left( x-z_i\right)
, 
\end{equation}
which shows that the topological structure of Chern density $\rho $ is
labeled by the Brouwer degrees and the Hopf index. The integration of $\rho
(x)$ 
\begin{equation}
\label{end}C_2=\int \rho (x)d^4x=\sum_{i=1}^m\eta _i\beta _i 
\end{equation}
is integer called Chern number which is a topological invariant$.$

The result (\ref{end}) suggests that the topological invariant is determined
only by the zero points of the vector field $\vec \phi ,$ i.e. the zeroes of
the spinning wavefunction $\psi .$ The topological indexes of zero points
are intrinsic properties of the wavefunction, so the topological invariant
given in (\ref{end}) is naturally inherent in the spinning system.

\section{Discussion}

Based on the gauge potential decomposition method, we established the gauge
potential decomposition theory in terms of the spinning wavefunction. One
finds that the $SU(2)$ gauge potential can be completely decomposed by the
wavefunction of spinning system. By making use of the $\phi $-mapping
theory, we reveal that the second Chern class is naturally inherent in
spinning system, which is only associated with the intrinsic properties of
the spinning wavefunction, and labeled by the topological indexes of the
wavefunction. We believe that this intrinsic topological property is a
fundamental property of quantum system and is the source of many topological
effects in quantum system. This gives a new concept in topological quantum
mechanics.

\section*{Acknowledgments}

This work was supported by the National Natural Science Foundation of China.


\begin{references}
\bibitem{t1}  G. Morandi, {\it The Role of Topology in Classical and Quantum
Physics}, Lecture Note in Physics Vol. {\bf M7} (Springer-Verlag, Berlin,
1992).

\bibitem{jmp28}  Albert S.Schwarz, {\it Topology For Physicist} (Springer
Verlag Press 1994).

\bibitem{t2}  A. P. Balachandran, Found. Phys. {\bf 24}(4), 455 (1994).

\bibitem{inden}  S. Nash and S. Sen, {\it Topology and Geometry of Physicists%
} (Academic Pre. INC. London. 1983); M.W. Hirsch, {\it Differential Topology}
(Springer Verlag. New York 1976).

\bibitem{D1}  Y. S. Duan and M. L. Ge, Sci. Sinica {\bf 11}, 1072 (1979);
Y.S. Duan and X.H. Meng, J. Math. Phys.{\bf \ 34,} 1149 (1993);

\bibitem{D2}  Y.S. Duan and X.G. Lee, Helv. Phys. Acta. {\bf 68 }513 (1995);
Y.S. Duan, S. Li and G.H. Yang, Nucl. Phys. {\bf B514 },705(1998).

\bibitem{D3}  Y. S. Duan and X. H. Meng, Int. J. Engng. Sci{\it .}{\bf \ 31, 
}1173 (1993).

\bibitem{D4}  Y. S. Duan and J. C. Liu, in {\it Proceedings of Johns Hopkins
Workshop 11} (World Scientific, Singapore, 1988).

\bibitem{D5}  Y. S. Duan and S. L. Zhang, Int. J. Engng. Sci{\it .} {\bf 28,}
689 (1990); {\bf 29,} 153 (1991); {\bf 29}, 1593 (1991); {\bf 30,} 153
(1992); Y. S. Duan, S. L. Zhang, and S. S. Feng, J. Math. Phys{\it .} {\bf %
35, }1 (1994); Y. S. Duan, G. H. Yang, and Y. Jiang, Gen. Rel. Grav. {\bf 29}%
, 715 (1997).

\bibitem{Dz}  Y.S. Duan, H. Zhang and S. Li, Phys. Rev. {\bf B58}, 125(1998).

\bibitem{fujmp}  Y.S. Duan and L.B. Fu, J. Math. Phys. {\bf 39, }4343(1998).

\bibitem{jmp22}  C. Doran, D. Hestenes, F. Sommen and N.V.Acker, J. Math.
Phys. {\bf 34 }3642 (1993);

\bibitem{jmp19}  D. Hestenes and G. Sobczyk,\ {\it Clifford Algebra to
Geometric Calculus} (Reidel, Dordrecht 1984).

\bibitem{jmp25}  Eguchi, Gilkey and Hanson, {\it Gravitation, gauge theories
and differential geometry}, Phys. Rep. {\bf 66} 213 (1980).

\bibitem{jmp27}  S.S. Chern and J. Simons, Ann. Math. {\bf 99 }48 (1974).

\bibitem{jmp29}  H. Hopf,\ Math. Ann. {\bf 96} 209 (1929).
\end{references}
\end{document}